\documentclass[11pt]{article}
\usepackage[utf8]{inputenc}
\usepackage{amsmath, amssymb, enumerate, graphicx, fullpage, physics}

\usepackage{natbib}
\usepackage{sidecap}
\usepackage{titling}
\usepackage{setspace}
\usepackage[margin=1in]{geometry}
\usepackage{overpic, rotating, pbox}

\usepackage{palatino} 

\newcommand{\St}{\mathrm{St}}

\usepackage{xcolor}

\title{\fontsize{21}{21}{\textbf{An empirical mean-field model of symmetry-breaking in a turbulent wake\vspace{-0.25in}}}}

\author{\normalsize{Jared L. Callaham$^{1*}$, Georgios Rigas$^2$, Jean-Christophe Loiseau$^3$, and Steven L. Brunton$^1$}\\
	\footnotesize{$^1$ Department of Mechanical Engineering, University of Washington,
    Seattle, WA 98195, USA}\\
	\footnotesize{$^2$ Department of Aeronautics, Imperial College London, London SW7 2AZ, UK}\\
	\footnotesize{$^3$ Arts et M\'{e}tiers Institute of Technology, CNAM, DynFluid, HESAM Universit\'{e}, F-75013 Paris, France\vspace{-.075in}}
	}

\begin{document}
\date{}
\maketitle
\vspace{-.5in}
\begin{abstract}
This work develops a low-dimensional nonlinear stochastic model of symmetry-breaking coherent structures from experimental measurements of a turbulent axisymmetric bluff body wake.
Traditional model reduction methods decompose the field into a set of modes with fixed spatial support but time-varying amplitudes.
However, this fixed basis cannot resolve the mean flow deformation due to variable Reynolds stresses, a central feature of Stuart's nonlinear stability mechanism, without the further assumption of weakly nonlinear interactions.
Here, we introduce a parametric modal basis that depends on the instantaneous value of the unsteady aerodynamic center of pressure, which quantifies the degree to which the rotational symmetry of the wake is broken.
Thus, the modes naturally interpolate between the unstable symmetric state and the nonlinear equilibrium.
We estimate the modes from experimental measurements of the base pressure distribution by reducing the symmetry via phase alignment and averaging conditioned on the center of pressure.
The amplitude dependence of the symmetric mode deviates significantly from the polynomial scaling predicted by weakly nonlinear analysis, confirming that the parametric basis is crucial for capturing the effect of strongly nonlinear interactions.
We also introduce a second model term capturing axisymmetric fluctuations associated with the mean-field deformation.  
We then apply the Langevin regression system identification method to construct a stochastically forced nonlinear model for these two generalized mode coefficients.
The resulting model reproduces empirical power spectra and probability distributions, suggesting a path towards developing interpretable low-dimensional models of globally unstable turbulent flows from experimental measurements.
\end{abstract}

\section{Introduction}
\label{sec:intro}
Despite being nominally deterministic, turbulent flows are characterized by multiscale spatiotemporal chaos. 
Many traditional theoretical and experimental analyses have therefore relied on statistical descriptions~\citep{MoninYaglom1}.
However, it is now known that many inhomogeneous flows are dominated by energetic coherent structures at large scales and low frequencies relative to Kolmogorov's universal equilibrium range~\citep{Aubry1988jfm, Berger1990jfs, Holmes1996}.
By leveraging this intrinsic structure, reduced-order models of turbulent flows promise to advance engineering objectives in design, optimization, and control~\citep{Noack2011book, Benner2015siam, Brunton2015amr, Rowley2017}.
However, balancing accuracy and efficiency by simultaneously modeling the evolution of the large-scale structures while accounting for the effects of incoherent fluctuations has been notoriously challenging, especially in a non-invasive fashion that is suitable for experimental measurements. 

In many cases, turbulent coherent structures can be linked to laminar instability modes.
Wake flows are a canonical example of this phenomenon.
Flows past bluff bodies at low Reynolds number often exhibit stereotypical global instabilities such as von K\`arm\`an vortex streets or steady symmetry-breaking wake deflection~\citep{Berger1990jfs, Noack2003jfm, Meliga2009jfm, Grandemange2012pre}.
Qualitatively similar coherent structures appear to dominate these wakes well into the turbulent regime~\citep{Grandemange2013pof, Grandemange2013jfm, Rigas2014jfm}.
A number of recent studies have leveraged this observation in model-based active flow control schemes for drag reduction~\citep{Osth2014jfm, Brackston2016jfm, Barros2016jfm}.

Despite the existence of high-energy coherent structures, the multiscale nature of turbulence poses a significant modeling challenge.
Operator-based methods such as resolvent analysis~\citep{McKeon2010jfm, Pickering2021jfm} or the harmonic balance method~\citep{Rigas2020jfm} offer a powerful set of numerical tools, but are generally too computationally intensive to scale to high Reynolds numbers and complex geometries of practical interest.
Semi-empirical modeling methods, such as Galerkin projection of the Navier-Stokes equations onto a proper orthogonal decomposition basis~\citep{Aubry1988jfm, Holmes1996, Noack2003jfm}, can be used to derive a compact approximation to the dynamics in the form of coupled amplitude equations.
However, projection-based methods often are unstable without additional closure modeling and require information about the flow field that is difficult to obtain experimentally.
Both approaches are intrusive, requiring access to a high-fidelity numerical solver. 

In order to address the analytic challenges of modeling turbulent flows, data-driven methods have long played a role in modal analysis~\citep{Taira2017, Taira2020aiaa} and reduced-order modeling~\citep{Noack2011book, Rowley2017}.
Recent advances in machine learning have generated increased interest in data-driven methods for fluid dynamics~\citep{Brenner2019prf, BarSinai2019pnas, Brunton2020arfm, Raissi2020science}, including schemes for turbulence modeling~\citep{Ling2016jfm, Duraisamy2018arfm, Maulik2019jfm, Beetham2020prf} and forecasting~\citep{Vlachas2018prs, Novati2021nature}.
Despite the expressive power of modern machine learning methods, it is challenging to develop models that are robust, generalizable, and interpretable.
The sparse identification of nonlinear dynamics (SINDy) framework~\citep{Brunton2016pnas} has promise as a simple, interpretable low-dimensional modeling technique~\citep{Loiseau2017jfm, Loiseau2020tcfd, Deng2020jfm}.
SINDy can also leverage additional information including symmetries, conservation laws, and invariant manifold structure~\citep{Loiseau2017jfm, Loiseau2018,champion2020unified}, although its success has been limited to laminar flows with the exception of recent work modeling RANS closures~\citep{Beetham2020prf}.

To address multiscale dynamics, such as turbulence,~\citet{Boninsegna2018jcp} and~\citet{Callaham2021langevin} recently extended SINDy for stochastic systems.
In contrast to systems that can be treated by classical nonequilibrium statistical mechanics, there is no strict separation of scales between low-frequency coherent dynamics and turbulent fluctuations.
Progress in output-only system identification has also advanced our ability to approximate Langevin dynamics from experimental data~\citep{Honisch2011pre, Boujo2016prs, Boujo2019jfs, Sieber2020, Frishman2020prx, Bruckner2020prl, Schneider2020}, even if the model structure is \textit{a priori} unknown~\citep{Boninsegna2018jcp}.
Because these methods are purely empirical, they may extend traditional stochastic modeling beyond near-equilibrium systems, providing a useful approximation despite the absence of a strict scale separation.
For example, the Langevin regression approach to sparse system identification was developed to correct for finite-time effects arising in stochastic approximations of multiscale systems~\citep{Callaham2021langevin}.

\begin{figure}
\vspace{-.175in}
	\centering
	\begin{overpic}[width=0.95\linewidth]{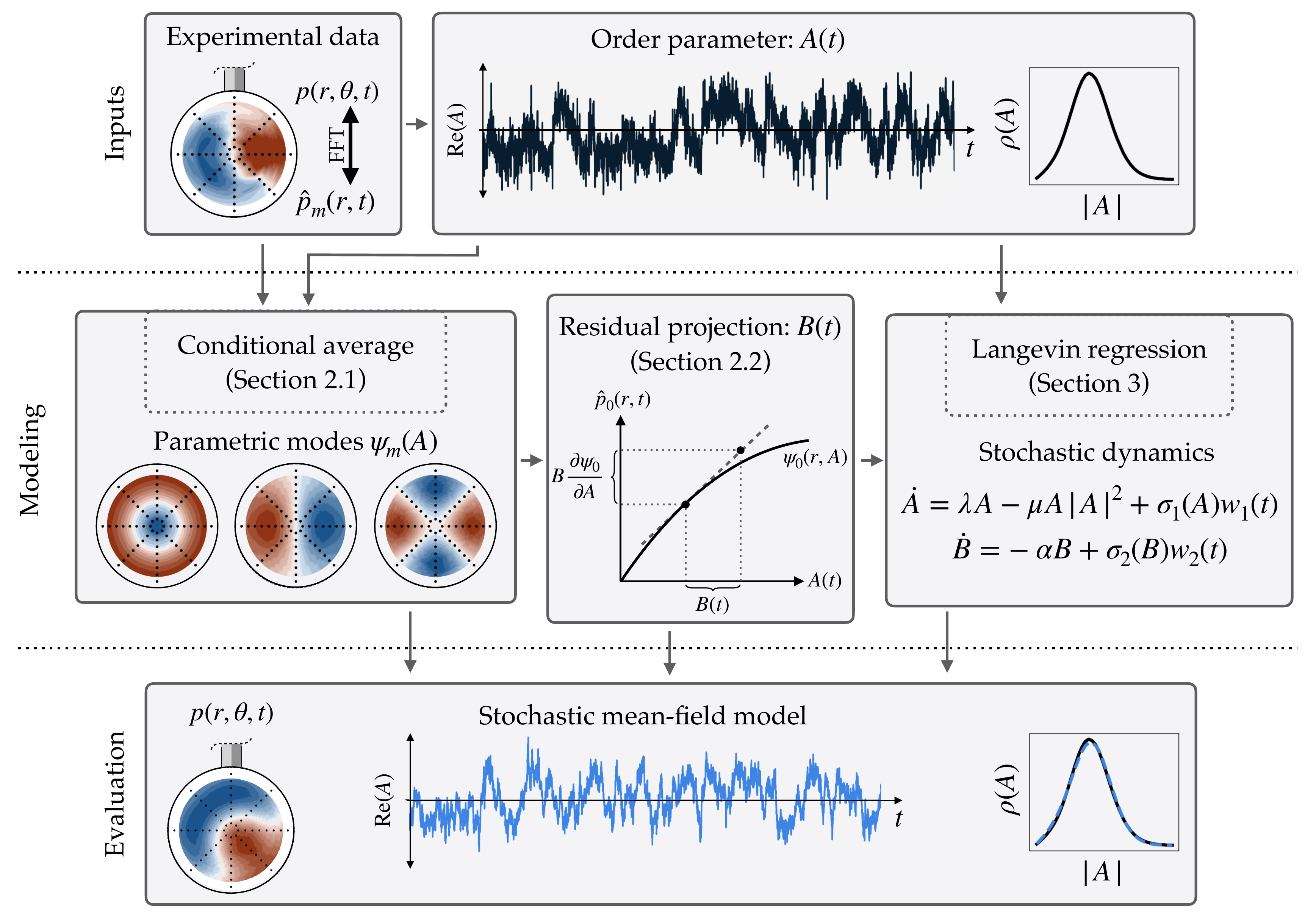}
	\end{overpic}
	\vspace{-.15in}
	\caption{Overview of the model development.
	Beginning with an order parameter $A(t)$ computed from the base pressure measurements, the spatial modes are computed by conditional averaging.
	These modes define an approximate slow manifold that captures the dominant anti-symmetric behavior.
	To fully resolve the symmetric fluctuations, we introduce a generalized shift mode with amplitude $B(t)$, defined by projection onto the tangent space of this manifold.
	Finally, we identify a nonlinear stochastic dynamical system model with Langevin regression.
	We can compare the statistics of the model to the experiment with Monte Carlo evaluation of the stochastic system.
	}
	\label{fig:flowchart}
	\vspace{-.15in}
\end{figure}
In this work we demonstrate a general data-driven modeling procedure (visualized in Figure~\ref{fig:flowchart}) that constructs stochastically forced nonlinear systems to model the evolution of coherent structures.
Specifically, we apply the recently proposed Langevin regression method~\citep{Callaham2021langevin} to identify an interpretable, low-order dynamical system that models the broadband turbulence as stochastic forcing of the deterministic symmetry-breaking behavior. 
We introduce both a parametric modal expansion based on conditionally averaging the experimental observations, and a stochastic dynamical model identified from the time series data.
The modal expansion reduces the physical field to a representation in terms of expansion coefficients (order parameters), while the stochastic model describes the evolution of these coefficients.
This approach is demonstrated for the reduced-order modeling of a fully turbulent wake behind an axisymmetric bluff body based on experimental measurements.
The resulting model is a stochastic Stuart-Landau equation similar to those proposed in previous studies of similar configurations~\citep{Rigas2015jfm, Boujo2019jfs, Herrmann2020}.
Monte Carlo evaluation of the Langevin system compares favorably to the statistics of the experimental pressure distribution.

Critically, the model does not include an axisymmetric ``bubble-pumping" modulation of the recirculation bubble, which does not appear in the laminar analysis but which has been proposed to explain a feature of the experimental power spectrum~\citep{Berger1990jfs}.
In our model, this peak is instead explained by a timescale of relaxation to the slow manifold in the dynamics of a secondary order parameter, suggesting that the bubble-pumping is not an independent physical mechanism but a result of perturbations to the dynamics of the dominant instability mode.

The paper is organized as follows.
Section~\ref{sec:background} provides an overview of relevant topics in nonlinear stability and mean-field theory, followed by a description of the flow configuration and experimental setup in Section~\ref{sec:wake}.
In Section~\ref{sec:mean-field} we introduce the mean-field model, including the order parameter and parametric modal basis.
We identify and evaluate a nonlinear Langevin model for the symmetry-breaking dynamics in Section~\ref{sec:dynamics}. 
Section~\ref{Sec:Discussion} concludes with a discussion.
Beyond the axisymmetric wake, these contributions represent a potential framework for developing descriptive low-dimensional models of turbulent flows from limited experimental measurements.

\section{Background on mean-field modeling}
\label{sec:background}
The energetic scales of many multiscale turbulent flows are dominated by semi-regular coherent structures whose spatial and dynamical features can often be qualitatively linked to laminar instability modes.
This motivates a ``triple decomposition'' approach to reduced-order modeling wherein the flow is conceptualized as the superposition of a steady mean, low-frequency coherent structures, and high-frequency, apparently stochastic forcing~\citep{Hussain1981}.
In this section we give a brief overview of some topics in low-dimensional modeling of turbulent flows with relevance to the axisymmetric wake.

\subsection{Weakly nonlinear analysis}
Fluid flows typically exhibit increasingly complex behavior with variation of a suitably-defined control parameter, such as the Reynolds number for incompressible flows.
One extreme is steady laminar flow with the same symmetries as the underlying geometry, while the other is fully-developed turbulence with symmetries only in the statistical sense.
The route between steady laminar flow and broadband turbulence is complex and depends on the flow configuration, but can involve general mechanisms, including non-normal energy growth and linear instability~\citep{Schmid2001book}.
Flows with reflectional or rotational symmetry (e.g. bluff body wakes or Taylor-Couette flow) often undergo global bifurcations leading to spatiotemporal symmetry-breaking~\citep{Cross1993, Meliga2009jfm, Hohenberg2015}.
Near the threshold of bifurcation, these flows can be modeled with low-dimensional weakly nonlinear dynamical systems describing the evolution of the unstable modes.
As the Reynolds number is increased, successive bifurcations eventually result in a breakdown into multiscale turbulence that cannot be described exactly with low-order dynamics.
Nevertheless, the dominant laminar instability modes often persist as large-scale coherent structures. 
Developing approximate reduced-order models for these coherent structures analogous to the weakly nonlinear amplitude equations is a long-standing challenge, promising to extend the analysis of laminar flows to the strongly nonlinear turbulence ubiquitous in engineering applications.

Nonlinear stability theory is a cornerstone of our understanding of globally unstable flows, both laminar and turbulent.
Linear stability analysis relies on the assumption of infinitesimal perturbations to the base flow, but predicts exponential energy growth for unstable modes.
In reality, the energy tends to saturate due to nonlinearities once the fluctuations reach finite amplitude.
One such interaction is a stabilizing feedback loop wherein the mean flow deformation due to self-interaction of the perturbations reduces the instantaneous growth rate of the instability mode~\citep{Stuart1958jfm, Landau1959book}.
This line of reasoning leads to the cubic Stuart-Landau equation governing the evolution of the instability mode amplitude $A(t)$:
\begin{equation}
\label{eq:stuart-landau}
    \dv{A}{t} = \left( \lambda - \mu |A|^2\right) A.
\end{equation}
The term in parentheses is the effective eigenvalue of the instability mode, as modified by the mean flow deformation.
When the real part of $\lambda$ is positive, small perturbations grow exponentially until the instantaneous growth rate reaches a balance with $\mu|A|^2$.

Similar amplitude equations can be derived from a rigorous asymptotic expansion close to the threshold of instability~\citep{Sipp2007jfm, Meliga2009jfm}, or via a manifold reduction of an empirical Galerkin model~\citep{Noack2003jfm}.
Both approaches resolve the nonlinear stability mechanism by introducing a basis of spatial modes that approximately spans the mean flow deformation.
On the other hand, an arbitrary deformation can be treated directly with the numerical self-consistent modeling method~\citep{ManticLugo2014prl, ManticLugo2016jfm}, although this does not lead to an amplitude equation.
While these methods have primarily been applied in laminar regimes, their broad applicability suggests a mechanism by which coherent structures related to linear instability modes might come to dominate turbulent flows.

\subsection{Mean-field theory of symmetry-breaking transitions}
\label{sec:background-thermo}
The Stuart-Landau nonlinear stability theory summarized above is typically derived as an asymptotic expansion in multiple timescales.
It therefore falls into the category of weakly nonlinear analysis and is only strictly applicable near the threshold of instability, although the effect is generally understood to persist much beyond the asymptotic regime in many cases~\citep{Noack2003jfm, Luchtenburg2009jfm, ManticLugo2014prl}.
However, Landau also considered another limiting case of dynamics: symmetry-breaking phase transitions of systems in thermodynamic equilibrium.
Here we briefly introduce the key features of this theory and in particular give the result that near-equilibrium dynamics can be described by an equation similar to the Stuart-Landau model~\eqref{eq:stuart-landau}. 
Turbulence is both strongly nonlinear and far from thermodynamic equilibrium, but the observation that both limiting regimes can be described with similar equations motivates the development of phenomenological Stuart-Landau-type models for symmetry-breaking behavior in the turbulent axisymmetric wake.
For a thorough introduction of mean-field theory and connection to weakly nonlinear analysis, see for example~\citet{Hohenberg2015}.

A generic system in thermodynamic equilibrium that undergoes a continuous symmetry-breaking phase transition at critical temperature $T_c$ (or temperature analogue such as Reynolds number) has an effective potential $\Phi(T, A)$ which we assume can be expanded in the magnitude of a (generally complex) order parameter $A$.
A canonical example of this is the Ising model, in which the statistically symmetric disorder of the high-temperature system is broken in a phase transition to a ferromagnetic state below a critical temperature.
Based on physical symmetries, the effective potential can be expanded as:
\begin{equation}
\label{eq:effective-potential}
    \Phi(T) = \Phi_0(T) + \Phi_1(T) |A|^2 + \Phi_2(T) |A|^4 + \dots.
\end{equation}
This order parameter, which here we assume to be small when suitably nondimensionalized, quantifies the degree of symmetry-breaking in the system.
The equilibrium condition $A_*$ at a given temperature is determined by the minimum free energy with respect to $|A|$.  To leading order,
$
    A_*(T) = \sqrt { -\Phi_1(T) / 2\Phi_2(T) }.
$

For the high Reynolds number axisymmetric wake studied in this work, the unsteady aerodynamic center of pressure serves as an order parameter capturing the symmetry-breaking wake deflection.
When nondimensionalized by the body diameter, its value is small even far from the bifurcation with mean value $\overline{A} \approx 0.032$.
The system is unsteady but statistically stationary.

In the thermodynamic perspective, the instantaneous field is disturbed from the minimum-potential state by broadband turbulence.
We model this as near-equilibrium thermal fluctuations in overdamped Langevin dynamics~\citep{Risken1996book}:
\begin{equation}
    \dv{A}{t} = -\nabla \Phi(A) + \Sigma(A) w(t),
\end{equation}
where $w(t)$ is a white noise process and $\Sigma$ is the diffusion function.
Expanding $\Phi(A)$ in $|A|$ to third order, the Langevin model would take the form
\begin{equation}
\label{eq:A-dynamics}
    \dv{A}{t} = \lambda A - \mu A |A|^2 + \Sigma(A) w(t),
\end{equation}
where $\lambda$ and $\mu$ are constants with positive real part.
This is identical to the Stuart-Landau equation \eqref{eq:stuart-landau} forced by white noise, although this argument is based only on symmetry and the scale of $|A|$ and doesn't assume proximity to a bifurcation.
In general the diffusion may be state-dependent; we identify a particular functional form with the Langevin regression as described in Section~\ref{sec:dynamics}.

\section{The turbulent axisymmetric wake}
\label{sec:wake}

\begin{figure}
	\centering
	\begin{overpic}[width=0.975\linewidth]{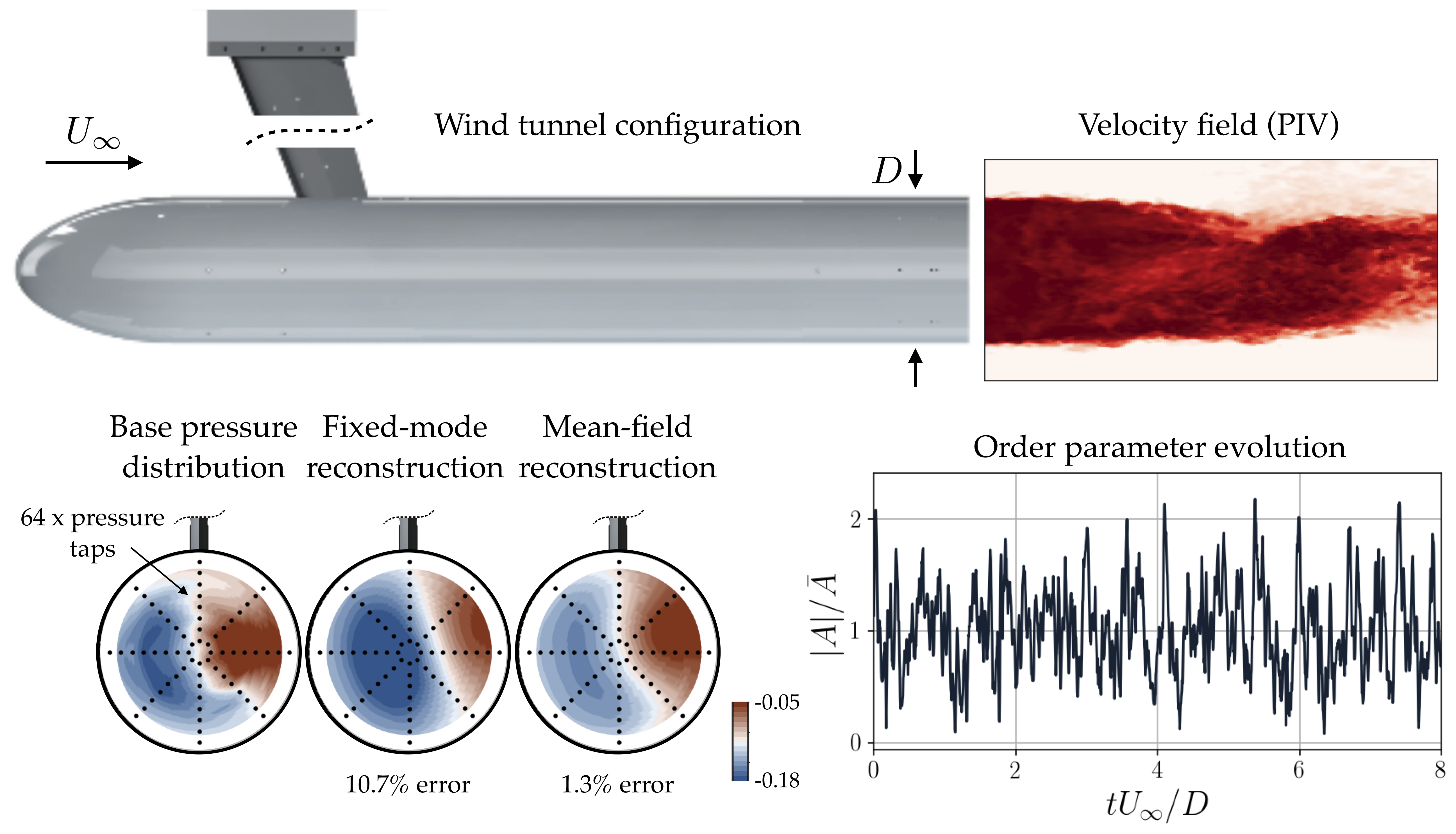}
	\end{overpic}
	\caption{Exerimental configuration for the axisymmetric wake (top).
	The signature of the dominant symmetry-breaking instability is captured by the complex order parameter defined in~\eqref{eq:order-parameter} (bottom right).
	The coherent fluctuations in the base pressure distribution can be approximated from the order parameter with the mean-field model (bottom left).
	The proposed model significantly improves the reconstruction over a traditional fixed-mode decomposition.}
	\label{fig:wake-config}
\end{figure}

The turbulent wake behind an axisymmetric bluff body, shown in Figure~\ref{fig:wake-config}, is a canonical flow that illustrates many of the modeling challenges outlined above.
Flow with free-stream velocity $U_\infty$, pressure $p_\infty$, density $\rho$, and kinematic viscosity $\nu$ is deflected around a cylindrical blunt-nosed body with diameter $D$.

The spatiotemporal dynamics of the wake are determined by the Reynolds number $\Re = DU_\infty/\nu$.
At very low Reynolds number, the wake is steady, axisymmetric and laminar.
The flow undergoes two bifurcations, becoming linearly unstable to a steady symmetry-breaking mode ($\Re_c^1 \approx 424$) and a second pair of unsteady vortex-shedding modes ($\Re_c^2 \approx 605$), both with azimuthal wavenumber $m=\pm 1$~\citep{Rigas2016ctr}.
By approximating these as a single codimension-2 bifurcation, the weakly supercritical flow can be approximated with an asymptotic expansion and normal form dynamics~\citep{Meliga2009jfm}.

These instability modes continue to dominate the coherent part of the flow even in fully developed turbulence, as shown for the present experimental data at $\Re \approx 2 \times 10^5$ by~\citet{Rigas2014jfm}.
The base pressure distribution is measured from a set of 64 equally spaced pressure taps from which a time series of $8.9 \times 10^6$ measurements are sampled at 225 Hz; further details are given in~\citet{Rigas2014jfm, Rigas2015jfm}.

Even though the time series of measurements appears strongly stochastic at this Reynolds number, the flow is characterized by semi-regular energetic structures, including vortex shedding and a symmetry-breaking wake deflection~\citep{Berger1990jfs, Grandemange2013jfm, Rigas2014jfm}.
These structures can be directly linked to the instability modes and weakly nonlinear dynamics of the corresponding laminar flow~\citep{Meliga2009jfm, Rigas2017jfm}.

Although the vortex shedding is dynamically important in the wake, it is only weakly observable from the pressure signal on the bluff body itself.
This also suggests that the vortex shedding is potentially less important to practical drag reduction than the symmetry-breaking wake deflection.
In this work we therefore restrict our attention to the steady symmetry-breaking instability and seek to describe its dynamics in terms of Stuart-Landau nonlinear stability theory.

The wake deflection in this flow is particularly important, as it represents generic symmetry-breaking behavior that occurs in a wide variety of three-dimensional bluff body wakes~\citep{Meliga2009jfm, Grandemange2012pre, Grandemange2013pof, Boujo2019jfs}.
Moreover, this symmetry-breaking is associated with increased pressure drag, making it the target of a variety of active flow control investigations~\citep{Osth2014jfm, Barros2016jfm, Brackston2016jfm, Rigas2017jfm, Herrmann2020}.
In several of these studies, reduced-order models have played a key role in designing and understanding the actuated system.

The pressure distribution on the base of the body is a convenient and experimentally accessible proxy for the coherent structures in the wake, since these limited observations can be clearly connected to previously observed wake structures based on symmetries and spectral energy content~\citep{Rigas2014jfm}.
The radial coordinate of the unsteady aerodynamic center of pressure is a natural order parameter for the degree of symmetry-breaking.

The instability mode responsible for the steady wake deflection in the laminar flow stabilizes at finite amplitude due to mean flow deformation~\citep{Meliga2009jfm}.
The turbulent wake exhibits similar behavior; \citet{Rigas2015jfm} and \citet{Boujo2019jfs} have modeled the aerodynamic center of pressure, closely related to the leading Fourier amplitude, with stochastic Stuart-Landau equations.
However, while the relationship between the amplitude equations and spatial mean flow deformation is clear for the weakly nonlinear laminar case, it has been unexplored in the phenomenological models developed for turbulent flows.

Standard model reduction methods decompose the field into a set of spatial modes~\citep{Taira2017,Taira2020aiaa} with coefficients whose time evolution is governed by the amplitude equations.
Within this framework, one way to resolve mean flow deformation is with the addition of a spatial mode parameterizing the difference between the unstable steady state and the mean flow.
This additional mode, often referred to as a \textit{shift mode}, can either be derived empirically~\citep{Noack2003jfm} or as a natural product of an asymptotic expansion~\citep{Sipp2007jfm}.
In either case, the assumption of weakly nonlinear interactions implies polynomial scaling of the deformation amplitude with respect to the amplitude of the dominant instability.
This approach has proven successful in low-dimensional models of a variety of laminar flows~\citep{Meliga2009jfm, Luchtenburg2009jfm, Loiseau2017jfm, Deng2020jfm}.
However, in strongly nonlinear turbulent flows this fixed modal basis cannot generally be expected to resolve the mean flow deformation.

\section{Parametric modal expansion}
\label{sec:mean-field}

In order to describe the symmetry-breaking and associated mean-field deformation, we model the evolution of the base pressure distribution $p(r, \theta, t)$, where $r$ and $\theta$ are polar coordinates on the circular bluff body base.
Although the pressure is not a dynamic variable in incompressible flow, the base pressure can be used as an experimentally accessible proxy for the energetic coherent structures in the wake~\citep{Rigas2014jfm}.
In this section we define an order parameter for the symmetry-breaking behavior and develop a parametric modal expansion that captures the correlated spatial fields.
Based on the symmetry of the flow, we expand the pressure field with a Fourier series and identify modes with a phase-aligned conditional average.
In contrast to fixed modal bases, such as the proper orthogonal decomposition, this approach allows the fields to naturally deform with amplitude.
We find that while the antisymmetric part of the field can be closely approximated with fixed spatial modes, the axisymmetric component deforms significantly and is not consistent with the scaling implied by weakly nonlinear analysis.

Let $p^0(r)$ be the pressure associated with the unstable axisymmetric steady state, which is unknown and experimentally inaccessible, and $\bar{p}(r)$ be the temporal mean field estimated from the pressure taps.
The mean flow is not necessarily a solution of the steady-state Navier-Stokes equations, but rather of the Reynolds-averaged equations.
The instantaneous fluctuations are determined with respect to this mean: $p'(r, \theta, t) = p(r, \theta, t) - \bar{p}(r)$.
The self-interaction of the velocity fluctuations associated with $p'$ deforms the unstable steady state to the mean flow via the Reynolds stresses.

The pressure field can be decomposed with a Fourier expansion in the azimuthal direction:
\begin{equation}
\label{eq:fourier-series}
p(r, \theta, t) = \sum_{m=-\infty}^\infty \hat{p}_m(r, t) e^{i m \theta}.
\end{equation}
Both the unstable steady state and the mean of the turbulent flow are axially symmetric ($m=0$), while the dominant instability modes are antisymmetric~\citep{Meliga2009jfm}.

Following~\citet{Rigas2015jfm} and \citet{Boujo2016prs}, we define the unsteady aerodynamic center of pressure on the bluff body base as a complex-valued order parameter $A(t)$:
\begin{equation}
\label{eq:order-parameter}
    A(t) = \frac{1}{\int p(r, \theta, t) r \dd r \dd \theta} \int p(r, \theta, t) r e^{i \theta} \dd r \dd \theta.
\end{equation}
The amplitude $|A(t)|$ is a measure of the degree of asymmetry in the wake, while its phase gives the azimuthal orientation of the wake deflection.
In practice we approximate this integral with a Riemann sum over the 64 pressure sensor locations.

\subsection{Coherent fields via conditional averaging}
\label{sec:mean-field-avg}
The order parameter amplitude $|A(t)|$ quantifies both the degree of asymmetry in the flow and the instantaneous strength of the coherent fluctuations, which are in turn responsible for the axisymmetric mean flow deformation and nonlinear equilibrium of the instability mode.
The mean field $\bar{p}$ and steady state $p^0$ are therefore associated with the mean amplitude $\bar{A} \equiv \overline{|A(t)|}$ and $A = 0$, respectively, although both fields are themselves axisymmetric.
Similarly, some instantaneous amplitude $|A(t)|$ between $0$ and $\bar{A}$ is associated with an axisymmetric field interpolating between $p^0(r)$ and $\bar{p}(r)$, though the amplitude itself only directly represents antisymmetric fluctuations.
In other words, the part of the instantaneous $m=0$ field that resolves the Stuart-Landau deformation mechanism is a direct function of the order parameter.
More broadly, we expect that the part of the field that is coherent with the order parameter can be revealed with an average conditioned on its amplitude.

This perspective suggests that the part of the field that is coherent with the symmetry-breaking might be approximated with a parametric modal decomposition:
\begin{equation}
\label{eq:modal-series}
\hat{p}_m(r, t) \approx \psi_m(r, A(t)).
\end{equation}
In contrast to a fixed space-time decomposition, such as proper orthogonal decomposition or dynamic mode decomposition~\citep{Schmid2010}, the spatial modes can deform according to the value of the instantaneous order parameter.
Once the modes $\psi_m(r, A)$ are known, such a representation reduces the field to a function of this single complex degree of freedom.

For example, if the self-interaction of the fluctuations are assumed to be weakly nonlinear, the axisymmetric component takes the particular polynomial form
\begin{equation}
\label{eq:shift-mode}
    \psi_0(r, A) = p^0(r) + |A(t)|^2 p_\Delta(r) + \mathcal{O}(|A|^4).
\end{equation}
In a numerical setting, the ``shift mode" $p_\Delta(r)$ can be determined either through an asymptotic expansion about the unstable base flow~\citep{Sipp2007jfm, Meliga2009jfm}, or from the difference between the steady-state and mean flows~\citep{Noack2003jfm}.

\begin{figure}
	\centering
	\begin{overpic}[width=0.95\linewidth]{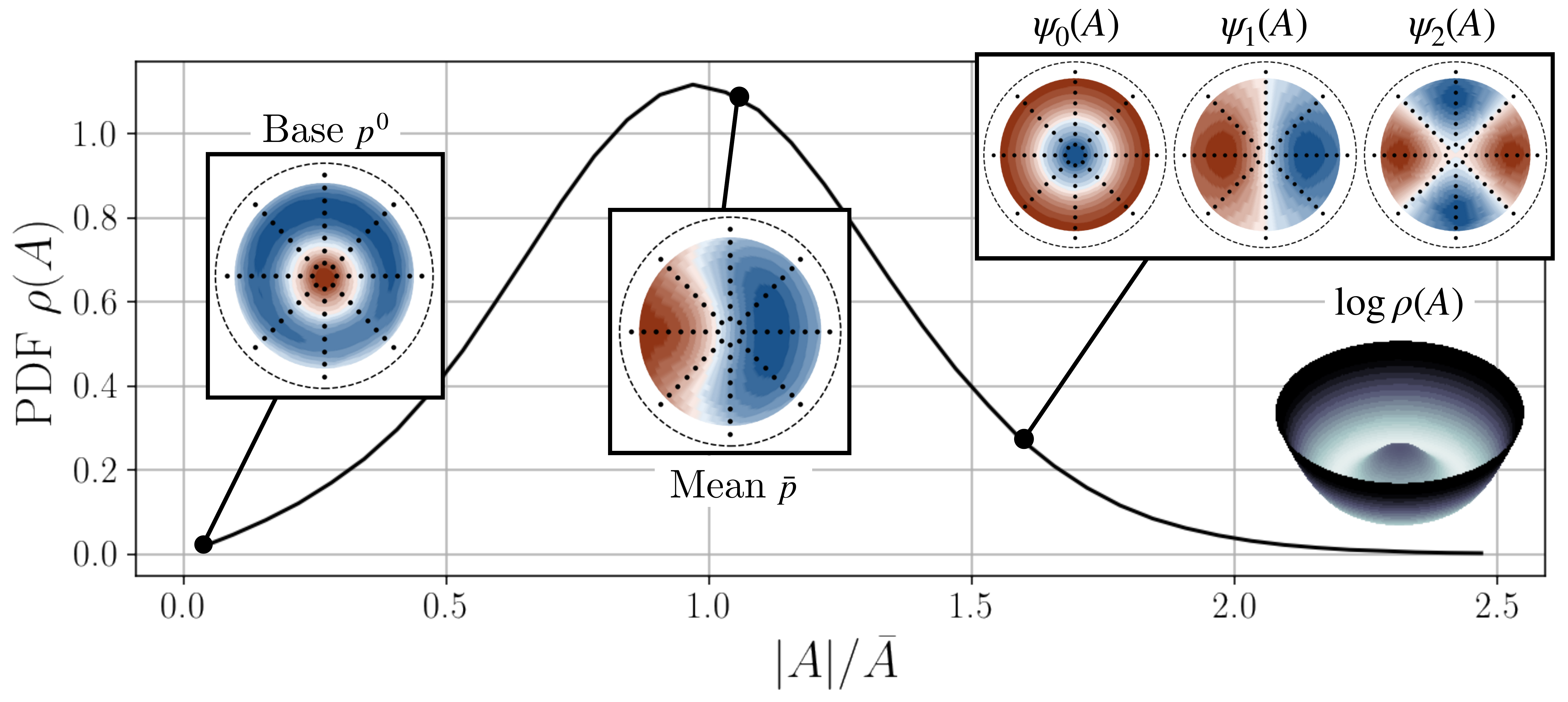}
	\end{overpic}
	\caption{Illustration of phase-aligned conditional averaging with respect to the radial center-of-pressure location $|A|$.
	We assume that the average at small amplitudes is representative of the pressure distribution associated with the unstable steady state $\boldsymbol{q}^B$.
	The field at any other point can be approximated with spline interpolation, allowing us to explore the amplitude dependence of the coherent fields.
	Also shown for reference is the unconditional phase-aligned mean field and the two-dimensional log-probability distribution, which is roughly analogous to a potential field.}
	\label{fig:cond-avg}
\end{figure}

Experimentally, neither of these approaches is viable, since the unstable steady state is typically unavailable.
Instead, we propose identifying the parametric modes with a phase-aligned conditional average on the order parameter, visualized in Figure~\ref{fig:cond-avg}.
The phase alignment reduces the symmetry of the fields; without this step all asymmetry would vanish on averaging.
On the other hand, the conditional average captures the natural variation of the field with the order parameter amplitude without any assumptions on the functional form of the $A$-dependence.

Beginning with the Fourier decomposition~\eqref{eq:fourier-series}, we compute the order parameter in amplitude-phase representation $A = |A|e^{i\phi}$.
The phase of the order parameter can then be removed from each field:
\begin{equation}
    p'(r, \theta, t) = \sum_m \hat{p}_m(r, t) e^{im (\theta - \phi(t))} = \sum_m \hat{p}'_m(r, t) e^{im\theta}.
\end{equation}
We divide the space of observed order parameter amplitudes $|A|$ into histogram bins centered on $A_i$ with width $2 \Delta A$.
For each wavenumber $m$ and histogram bin $i$, the radial component of $\psi_m(r, A_i)$ is approximated with
\begin{equation}
\label{eq:cond-avg}
\psi_m (r, |A_i|) = \left \langle \hat{p}'_m(r, t)  \bigg |  \big | |A(t)| - |A_i| \big | < \Delta A \right \rangle.
\end{equation}
We also estimate the base field $p^0(r)$ as the conditional mean at $m=0$ for the smallest histogram, i.e. over fields for which $|A(t)| < \Delta A$.

\begin{figure}
	\centering
	\begin{overpic}[width=\linewidth]{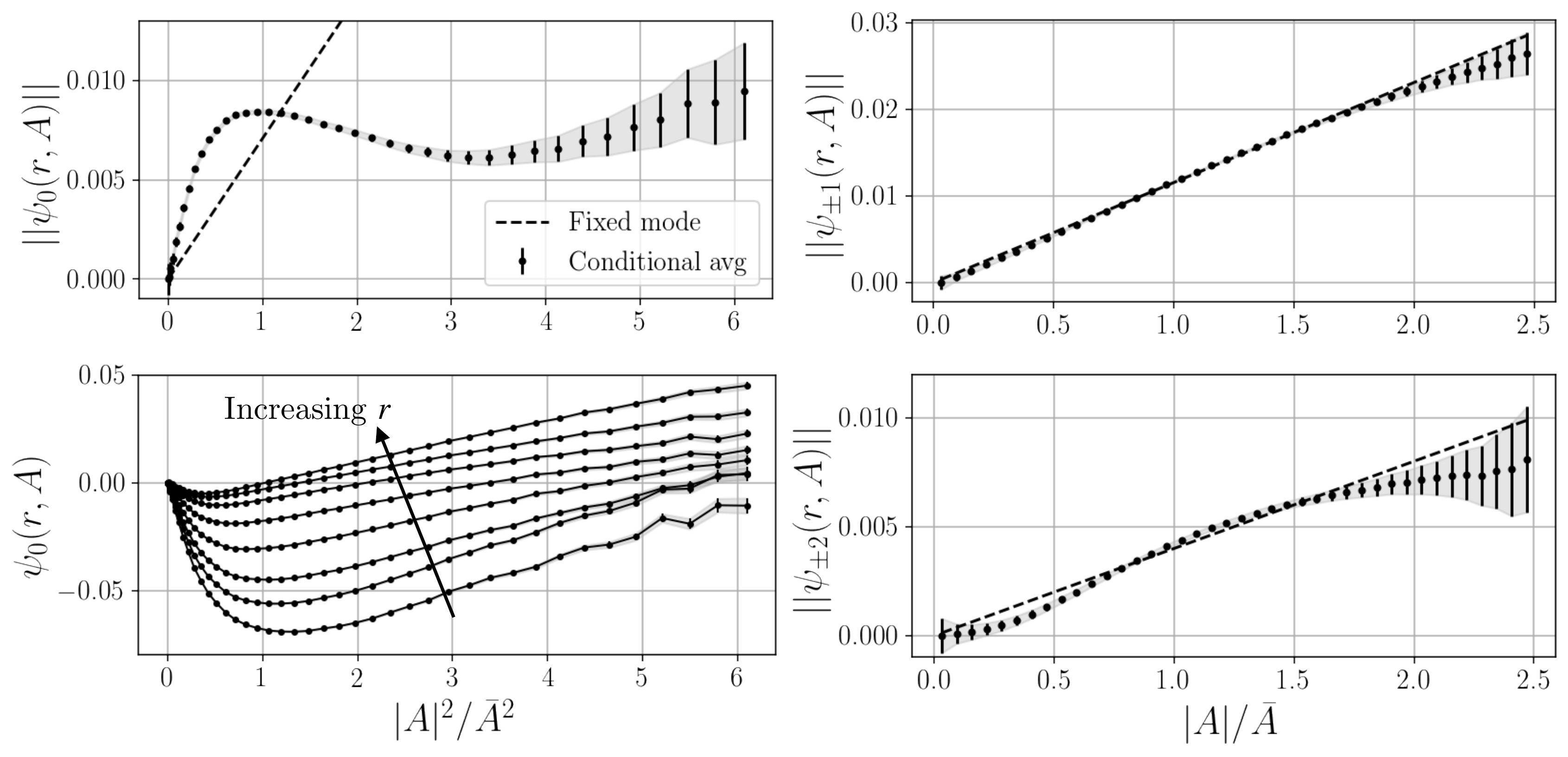}
	\put(9.3,44.5){a)}
	\put(58.8,44.5){b)}
	\put(9.3,22){c)}
	\put(58.8,22){d)}
	\end{overpic}
	\caption{The phase-aligned conditional average of the experimental pressure fields gives the coherent fields as a function of instability amplitude $|A|$ for various azimuthal wavenumbers $m$.
	The deformation of the axisymmetric part of the field cannot be explained by a single fixed mode or weakly nonlinear scaling (a), even though the deformation has a smooth dependence on the amplitude (c).
	On the other hand, the symmetry-breaking fields at $m=\pm 1, \pm 2$ are consistent with the fixed-mode assumption of Stuart-Landau theory (b, d).
	}
	\label{fig:deformation}
\end{figure}
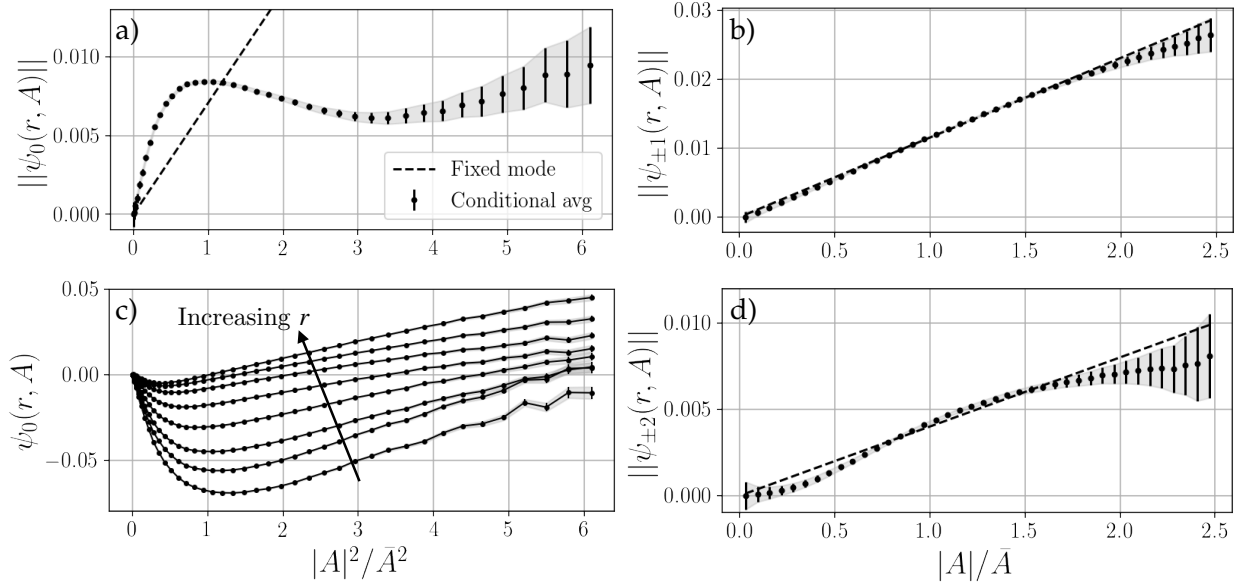

The procedure estimates the part of the field at each wavenumber that is correlated with the order parameter amplitude $|A(t)|$.
A continuous estimate of the modes can then be constructed with a spline interpoation of the conditional averages $\psi_m(r, |A_i|)$.
Figure~\ref{fig:deformation} shows the radially integrated modes (a, b, d), along with the axisymmetric mode at each radial sensor location (c).
The conditional average is compared to a fixed mode approximation where the spatial structure is fixed at its value at the mean amplitude $\bar{A}$.
We draw several conclusions from the amplitude scaling of these modes.

First, the axisymmetric field  at $m=0$ cannot be described by a fixed mode.
The weakly nonlinear scaling $\psi_0 \sim |A|^2$ clearly does not hold for typical amplitudes in this case, as shown by Figure~\ref{fig:deformation}a.
Figure~\ref{fig:deformation}c also shows the value of the $m=0$ field at each of the 8 radial sensor locations as a function of $|A|^2$; a single fixed mode cannot explain this behavior even if its integrated amplitude is a complicated function of $|A|$.
This indicates that the nonlinear axisymmetric  self-interaction and higher antisymmetric harmonics play a significant role in altering the spatial structure of the axisymmetric deformation as the fluctuation amplitude changes.

However, to a good approximation both the $m=\pm 1$ and $m=\pm 2$ Fourier components can be reasonably well-described by a single fixed mode with linear dependence on $|A|$, as shown in Figure~\ref{fig:deformation}b, d.
This is consistent with the typical assumption of the Stuart-Landau theory that the instability is a fixed eigenmode of the linear operator with variable eigenvalue.
Higher wavenumbers show weak coherence with the order parameter; we therefore truncate the reconstruction at $|m|=2$.

\begin{figure}
	\centering
	\begin{overpic}[width=0.85\linewidth]{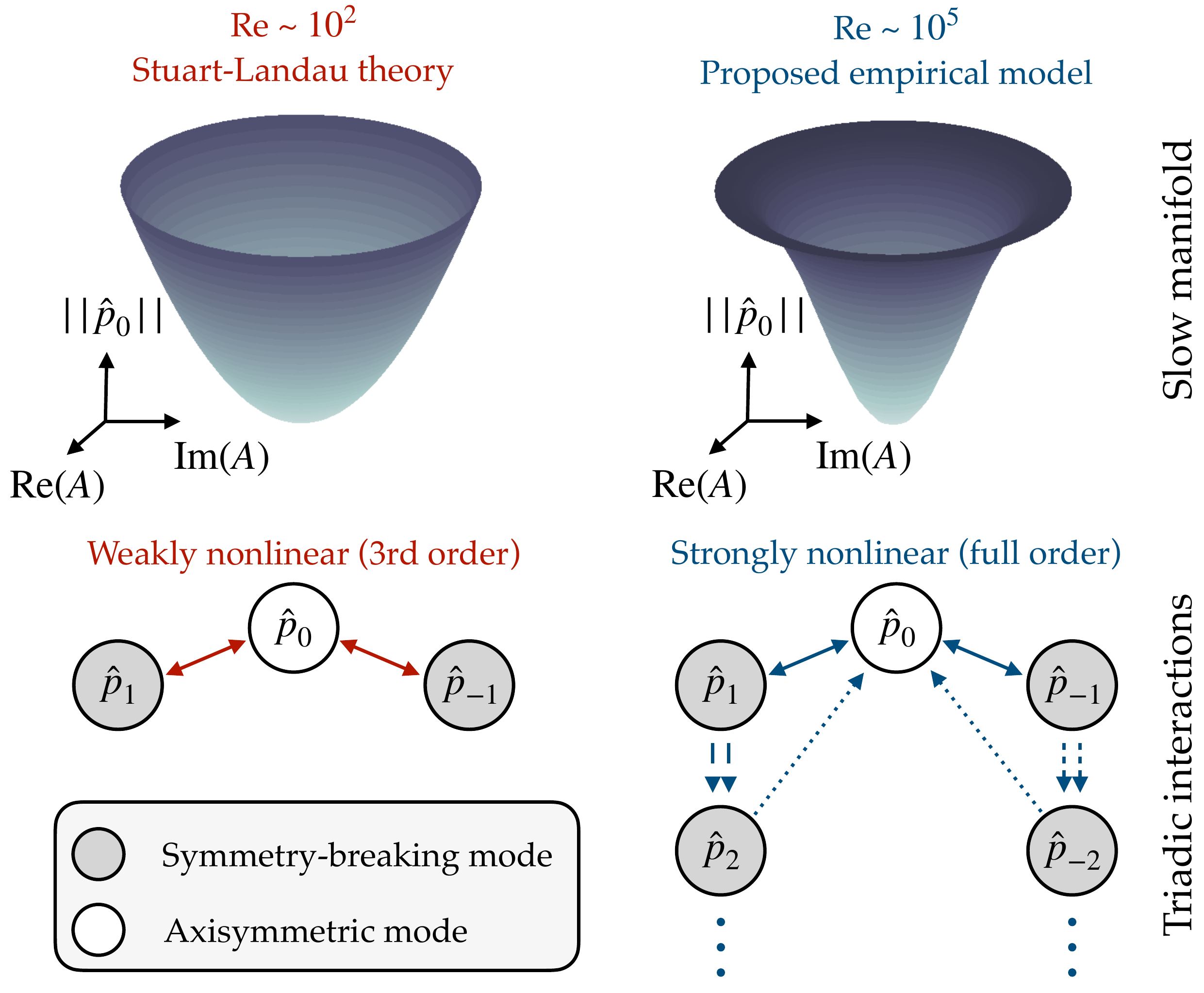}
	\end{overpic}
	\caption{ Visualization of approximate slow manifolds and the nonlinear interactions responsible for generating them. 
	Weakly nonlinear analyses of laminar flows typically neglect higher-order interactions for small fluctuations, leading to a parabolic manifold for the low-dimensional dynamics (left).
	Using conditional averaging, we show that these interactions are necessary to explain observations of the turbulent wake (right).
	Pairs of lines with similar styles indicate the structure of the leading-order triadic interactions.
	The physical interactions are between velocity components, but here we use the base pressure field as a proxy for coherent structures in the wake.}
	\label{fig:triadic-manifold}
\end{figure}

The conditional average in $A$ can be viewed as an empirical approximation of the slow manifold related to the symmetry-breaking behavior.
This manifold may be visualized by revolving the conditional deformation in Figure~\ref{fig:deformation} about the $A$-axis, as shown in Figure~\ref{fig:triadic-manifold}.
The weakly nonlinear scaling~\eqref{eq:shift-mode} generates the parabolic ``theoretical" manifold, while the parametric modes~\eqref{eq:cond-avg} generate its experimental counterpart.
The empirical manifold naturally accounts for nonlinear interactions at all orders, so this comparison reveals the importance of nonlinear self-interaction of the axisymmetric component and higher-order triadic interactions.

This conditional averaging approach can be viewed as an empirical approximation to the self-consistent mean field model~\citep{ManticLugo2014prl}, since it gives the fields at arbitrary fluctuation amplitudes without assuming a fixed spatial structure as in standard modal analyses.
Contrary to the self-consistent model, however, we do not neglect the influence of higher harmonics on the base flow.
Since the modes are derived directly from experimental data, they naturally account for higher-order interactions.

\begin{figure}
	\centering
	\begin{overpic}[width=0.9\linewidth]{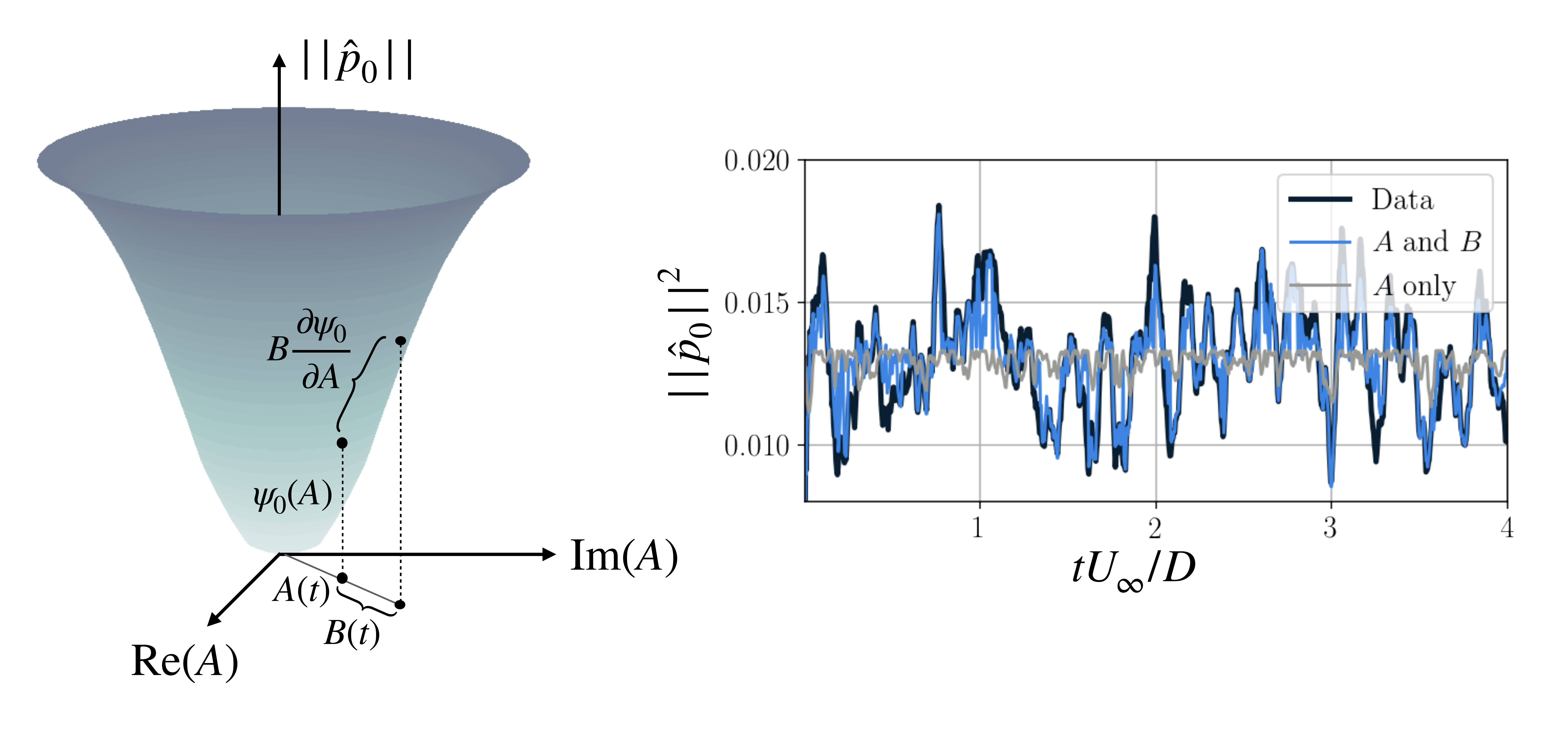}
	\end{overpic}
	\caption{Model for axisymmetric fluctuations.
	The conditional average on the order parameter $A$ defines a slow manifold (left), but including an additional degree of freedom to account for fluctuations significantly improves the resolution of the $m=0$ component (right).}
	\label{fig:gradient_mode}
\end{figure}

Finally, we reiterate the difference between the decomposition and the shift mode or weakly nonlinear results.
In these cases the fixed space-time decomposition and assumptions of weak nonlinearity lead to the polynomial amplitude scaling in~\eqref{eq:shift-mode}, which neglects self-interaction of the axisymmetric component as well as influences of higher harmonics.
On the other hand,~\eqref{eq:reconstruction} can resolve arbitrary deformations of the axisymmetric component due to these effects and does not constrain the functional form of the amplitude dependence.
As we will argue in Sec.~\ref{sec:dynamics}, this general parametric dependence does not necessarily invalidate the previously proposed cubic Stuart-Landau model for the evolution of the order parameter.

\subsection{Residual projection for mean flow modification}
Since the symmetry-breaking is linked to an instability mode, we expect that the antisymmetric modes will not significantly deform with amplitude, so that $\psi_m(r, |A|) \approx |A(t)| \psi_m(r)$ for $m \neq 0$.
However, the coherent part of the axisymmetric component is dominated by nonlinear deformations induced by the Reynolds stresses, so in general it may have a complicated amplitude dependence.

For weakly nonlinear laminar flows, an asymptotic analysis indicates that the axisymmetric deformations can be expressed directly as a function of the instability amplitude $|A|^2$, ``pinning" the state to the slow manifold~\citep{Sipp2007jfm, Meliga2009jfm}.
However, the structure of reduced-order models based on Galerkin projection onto a set of fixed spatial modes suggests that in general there may be some finite relaxation time to the equilibrium state determined by $|A|$~\citep{Noack2003jfm}.
For example,~\citet{Sieber2020} assume a stochastically forced form of the laminar model proposed by~\citet{Noack2003jfm} along with a fixed set of spatial modes in order to model a symmetry-breaking instability in a turbulent swirling jet.
However, the form of this model is fundamentally based on weakly nonlinear arguments, which do not necessarily maintain validity in fully developed turbulence, as demonstrated in Sec.~\ref{sec:mean-field-avg} for the scaling of the axisymmetric deformations.

In other words, the expansion~\eqref{eq:modal-series} defines an approximate slow manifold (as depicted in Figure~\ref{fig:triadic-manifold}), though we would like to avoid assuming the state always resides on this surface.
We address this by introducing an additional (real) degree of freedom $B(t)$ representing the difference between the axisymmetric field $\hat{p}_0(r, t)$ and the manifold field $\psi_0(r, |A(t)|^2)$.
If this difference is typically small, the axisymmetric field can be expressed as a linearization about $|A(t)|$:
\begin{equation}
\label{eq:taylor-expansion}
    \hat{p}_0(r, |A(t)|, B(t) ) \approx \psi_0(r, |A(t)|^2 ) + 2 |A| B \left.\pdv{\psi_0}{|A|^2}\right|_{|A(t)|^2}.
\end{equation}
Defining the new $m=0$ mode
\begin{equation}
    \psi_B(r, |A|) = \left.\pdv{\psi_0}{|A|^2}\right|_{|A(t)|^2},
\end{equation}
it is clear that this model for the axisymmetric part of the field is a generalization of the fixed ``shift mode" model proposed by~\citet{Noack2003jfm}.
In particular, if the axisymmetric field does have the weakly nonlinear scaling $\psi_0(r, |A|^2) = |A|^2 \psi_0(r)$ and the linearization is about the unstable fixed point $A = 0$, then the tangent field in~\eqref{eq:taylor-expansion} is just $\psi_0(r)$.
For models based on Galerkin projection it is more natural to define a single coefficient for each mode.
That is, in the augmented POD models introduced by~\citet{Noack2003jfm} and~\citet{Sieber2020}, $B$ is defined so that
\begin{equation}
    \hat{p}_0(r, B(t) ) \approx B(t) \psi_0(r).
\end{equation}
However, the proposed parametric expansion in the present work allows the model more flexibility to capture the natural variations of the flow, without assuming any scaling behavior.

The instantaneous value of this secondary order parameter $B(t)$ can be estimated by projecting the part of the axisymmetric field not correlated with $|A|$ onto the tangent space of the slow manifold:
\begin{equation}
    B (t) \approx \frac{ \int \left( \hat{p}_0(t) - \psi_0(t) \right) \psi_B(t) r \dd r}{\int \psi_B(t) \psi_B(t) r \dd r},
\end{equation}
where the explicit dependence on $r$ and $|A(t)|$ has been omitted everywhere for notational clarity.
This residual projection procedure is depicted in Figure~\ref{fig:gradient_mode}.
In practice, if the axisymmetric mode $\psi_0(r, |A|)$ is interpolated in $|A|$ with a spline, the tangent field $\psi_B(r, |A|)$ can be computed with a derivative of the spline at each radial position $r$.

We emphasize that the conditional averaging has already resolved the axisymmetric deformations associated with the Stuart-Landau nonlinear stability mechanism.
Additionally, this secondary order parameter $B(t)$ is not necessary to describe the symmetry-breaking behavior, but as we will show, it significantly improves the resolution of the model for the axisymmetric fluctuations.

With this modal expansion, the base pressure field can be reconstructed with
\begin{align}
\label{eq:reconstruction}
p(r, \theta, t) =  \underbrace{p^0(r) + \psi_0(r, |A(t)|) + B(t) \psi_B(r, |A(t)|) }_\textrm{axisymmetric} + \underbrace{|A(t)| \sum_m \psi_m(r) e^{im(\theta + \phi(t))} }_\textrm{antisymmetric}.
\end{align}
The full model consists of the unstable steady state $p^0$, the symmetric deformation $\psi_0$, the linearization of the manifold $\psi_B$, the instability mode at wavenumber $m=\pm 1$, and its harmonics at $|m| > 1$.
However, the model only has three degrees of freedom: the complex instability amplitude $A(t)$ and the real axisymmetric deviation from the manifold $B(t)$.
Moreover, all of these modes can be directly computed from experimental observations, along with parametric variation in $A$.
An example reconstructed field is shown in Figure~\ref{fig:wake-config}.

Figure~\ref{fig:gradient_mode} also compares the integrated amplitude of the axisymmetric $m=0$ component between the experimental pressure distribution (black), reconstruction from only the order parameter $A$ (grey), and including the projected residuals $B$ (blue).
The additional gradient mode $\psi_B$ and secondary order parameter $B$ significantly improves the resolution of axisymmetric fluctuations, indicating that the linearization~\eqref{eq:taylor-expansion} and additional degree of freedom are both necessary and sufficient to approximate the $m=0$ Fourier component.

Table~\ref{tab:error} compares the reconstruction error of the various modal expansions with the normalized error metric
\begin{equation}
    \label{eq:error-metric}
    \varepsilon(p, \tilde{p}) = \frac{ \iint \left( p(r, \theta) - \tilde{p}(r, \theta) \right)^2 r \dd r \dd \theta } { \iint p(r, \theta)^2 r \dd r \dd \theta }.
\end{equation}
A reconstruction based only on the instability mode $\psi_{\pm 1}$ is used as a baseline, while the other models account for both the $m=0$ fluctuations and the higher harmonic at $m=\pm 2$.
On average, the fixed shift mode approximation~\eqref{eq:shift-mode} does worse than disregarding the fluctuations altogether.
Allowing parametric variations in $\psi_0$ by interpolating the conditional average only improves marginally compared to disregarding unsteady axisymmetric fluctuations altogether, while the full mean-field model including $\psi_B$ reduces the error by approximately 42\%.
This suggests that the axisymmetric component of the field has dominant low-dimensional structure captured by $B$ that is related to, but linearly uncorrelated with, the symmetry-breaking.

\begin{table}[t]
\vspace{.15in}
\centering
 \begin{tabular}{||c ||c c c c||} 
\hline 
& $\psi_{\pm 1}$ only  & Shift mode & \shortstack{Parametric $\psi_0$\\$A$ only} & \shortstack{Full mean-field\\ $A$ and $B$}\\ [0.5ex] 
 \hline\hline
 Error $\varepsilon$ & 0.0274 & 0.0310 & 0.0267 & 0.0160 \\ 
 \hline
 Reduction in $\varepsilon$ & -- & -13 \% & 2\% & \textbf{42\%} \\
 \hline
\end{tabular}
\vspace{.15in}
\caption{Normalized reconstruction error~\eqref{eq:error-metric} of modal expansions along with percent error reduction compared to using only the $\psi_{\pm 1}$ field (disregarding axisymmetric fluctuations and $|m|>1$).}
\label{tab:error}
\end{table}

If the weakly nonlinear scaling held, we would expect that the shift mode would approximately resolve the axisymmetric deformations; this failure is consistent with the disagreement of amplitude scaling in Figure~\ref{fig:deformation}.
Furthermore, the significant reduction in approximation error with the additional degree of freedom $B$ suggests that the instantaneous realization of the axisymmetric field is often different from its averaged value on the ``slow manifold".
Based on these observations, we hypothesize that the amplitude scaling and mean-field relationships in fully developed turbulence are not simple extrapolations of the established laminar theory.

\section{Reduced-order model of symmetry-breaking}
\label{sec:dynamics}

In this section we apply Langevin regression to the generalized modal coefficients $A(t)$ and $B(t)$, approximating the incoherent turbulent fluctuations as Gaussian white noise.
We identify both the form and coefficients of the model from the experimental data with the Langevin regression method for identifying nonlinear stochastic systems of the form
\begin{equation}
    \dv{\boldsymbol{x}}{t} = \boldsymbol{f}(\boldsymbol{x}) + \boldsymbol{\sigma}(\boldsymbol{x}) \boldsymbol{w}(t),
\end{equation}
where $\boldsymbol{w}(t)$ is Gaussian white noise.

Reduced-order models of turbulent flows are generally challenging to construct for the same reason that the statistical perspective was so successful historically; signals are dominated by apparently random turbulent fluctuations.
The most common approach to dealing with the fluctuations is to introduce a closure model that approximately captures their effects, for instance via an eddy viscosity term to achieve the correct dissipation, without explicitly treating the fast scales~\citep{Aubry1988jfm, Holmes1996, Noack2008jnet}.
A major difficulty in treating the fast fluctuations directly is that there is not a strict enough separation of scales to apply the machinery of near-equilibrium statistical mechanics.
Although they cannot be strictly justified, in practice some approximations based on statistical mechanics have been successful.
For example, modeling coherent structure dynamics with a generalized Langevin equation assumes that the fluctuations are normally distributed and uncorrelated in time.
While Eulerian statistics are usually non-Gaussian, central limit theorem arguments suggest that both Lagrangian variables and integral quantities have near-normal distributions~\citep{MoninYaglom1, Kraichnan1989}.
Similarly, although the turbulent fluctuations are correlated in time, various modeling methods have nevertheless been successful by approximating them as white noise~\citep{Majda2001cpam, McKeon2010jfm, Rigas2015jfm}.
Alternatively, recent work has investigated the use of colored noise in linearized flow models~\citep{Zare2017jfm, Towne2021aiaa}, although classical statistical physics tools such as the Fokker-Planck equation cannot be applied in this case.

Another complication that may arise in Langevin equations is state-dependent diffusion.
This may seem like an artificial modification, since the turbulent fluctuations are expected to be locally isotropic, and therefore have amplitudes that are independent of the large-scale motions.
However,~\citet{Majda2001cpam} and \citet{Pradas2012} demonstrate with asymptotic multiscale analyses that state-dependent diffusion can arise naturally via quadratic nonlinearities, even when only the fast scales are forced by constant-amplitude, additive white noise.

In previous work the evolution of the order parameter magnitude $|A(t)|$ has been successfully modeled by a stochastic cubic amplitude equation, inspired by the weakly nonlinear normal form~\citep{Rigas2015jfm}.
However, the weakly nonlinear analysis is predicated on a fixed-mode decomposition, which is at odds with the proposed amplitude-dependent spatial modes in Section~\ref{sec:mean-field}.
Nevertheless, a dynamical model with similar structure can also arise from the mean-field theory of symmetry-breaking phase transitions as introduced in Section~\ref{sec:background-thermo}, which does not rely on fixed spatial modes or weak nonlinearity.

\subsection{Nonlinear stochastic system identification: Langevin regression}
Although this qualitative symmetry-based argument suggests the expected structure of the Langevin equations, we use the recently proposed Langevin regression method for identifying nonlinear stochastic models~\citep{Callaham2021langevin} to identify the drift and diffusion functions rather than presuppose this form.
This method, a stochastic variant of the sparse identification of nonlinear dynamics (SINDy) approach~\citep{Brunton2016pnas}, optimizes free parameters of a Langevin model via solutions of both the forward and adjoint Fokker-Planck equations.
The steady-state solution of the forward equation gives the steady-state probability distribution, while the adjoint solution gives the finite-time propagation statistics of the Langevin model~\citep{Risken1996book}.
Langevin regression simultaneously minimizes the discrepancy between the Fokker-Planck solutions and statistics computed from the experimental data.
This system identification approach does not require access to the governing equations and can be applied to arbitrary quantities of interest.

Moreover, this approach can be combined with the simple stepwise sparse regression (SSR) procedure for selecting a parsimonious, but maximally descriptive model from a set of candidates~\citep{Boninsegna2018jcp, Callaham2021langevin}.
This algorithm sequentially discards functions whose exclusion is associated with the smallest increase in cost function.
The Pareto-optimal model can then be chosen as the simplest model with a small cost function, as shown in figure~\ref{fig:ssr}.
Based on the discussion in Section~\ref{sec:background-thermo}, we use a library of polynomials consistent with the expected symmetries as candidate functions.
The drift and diffusion functions are approximated by sparse linear combinations of these functions.
For instance, the drift and diffusion libraries for $A$ are
\begin{subequations}
\begin{align}
    \mathbf{\Theta}_f^A(A) &= \begin{bmatrix} A & A|A|^2 & A|A|^4 & A|A|^6 \end{bmatrix} \\
    \mathbf{\Theta}_\sigma^A(A) &= \begin{bmatrix} 1 & |A| & |A|^2 \end{bmatrix}.
\end{align}
\end{subequations}
Then the approximate drift function is $f_A(A) = \mathbf{\Theta}_f^A(A) \boldsymbol{\xi}_f^A$, where $\boldsymbol{\xi}_f^A$ is a relatively sparse coefficient vector that identifies the library terms that are active in the dynamics.
We do not include the axisymmetric deviation amplitude $B$ because it is not associated with any symmetry-breaking in the flow and therefore would not appear in the effective potential~\eqref{eq:effective-potential}.
The rotational symmetry of the physical system implies that $\boldsymbol{\xi}_f^A$ should be purely real, but since Langevin regression is based on a least-squares method a small imaginary part will generally be retained.
For the sake of a minimum-complexity model we enforce that $\boldsymbol{\xi}_f^A$ is real, although allowing complex-valued coefficients does not noticeably change the results.

Likewise, the libraries for $B$ are
\begin{subequations}
\begin{align}
    \mathbf{\Theta}_f^B(B) &= \begin{bmatrix} B & B^2 & B^3 \end{bmatrix} \\
    \mathbf{\Theta}_\sigma^B(B) &= \begin{bmatrix} 1 & B & B^2 \end{bmatrix}.
\end{align}
\end{subequations}
Again, we do not expect strong coupling between the order parameters $A$ and $B$, since $B$ was defined primarily to capture fluctuations that were uncorrelated with $A$.

Decoupling the dynamics also reduces the maximum dimensionality of the regression problems from three dimensions to two, since the real and imaginary parts of $A$ must be treated separately.
Since Langevin regression relies on approximating statistical quantities with histograms and solving the Fokker-Planck equation on a discretized grid, it does not scale well to higher dimensions.
For problems with multiple coupled instability modes, histogram-free approaches such as Langevin inference~\citep{Bruckner2020prl, Frishman2020prx} may be more appropriate, although this does not enforce consistency with the steady-state probability distribution.

The two key parameters in this method are the finite sampling rate, which allows the fast turbulent fluctuations to appear approximately uncorrelated in the time series, and the weight of the Kullback-Leibler (KL) divergence between the empirical probability distribution and the steady-state solution of the Fokker-Planck equation.
The latter controls the relative weight factor of the forward and adjoint Fokker-Planck solutions in the objective function.
Some heuristics for selecting these parameters are described in~\citet{Callaham2021langevin}.
We choose a sampling rate 200 times slower than the experimental sampling rate and select the KL divergence weight so that the forward and adjoint Fokker-Planck solutions have approximately equal contributions in the optimization.
The values are $10^{-2}$ and $10^2$ for the $A$ and $B$ optimizations, respectively. 

\subsection{Model analysis and evaluation}

\begin{figure}
	\centering
	\begin{overpic}[width=0.85\linewidth]{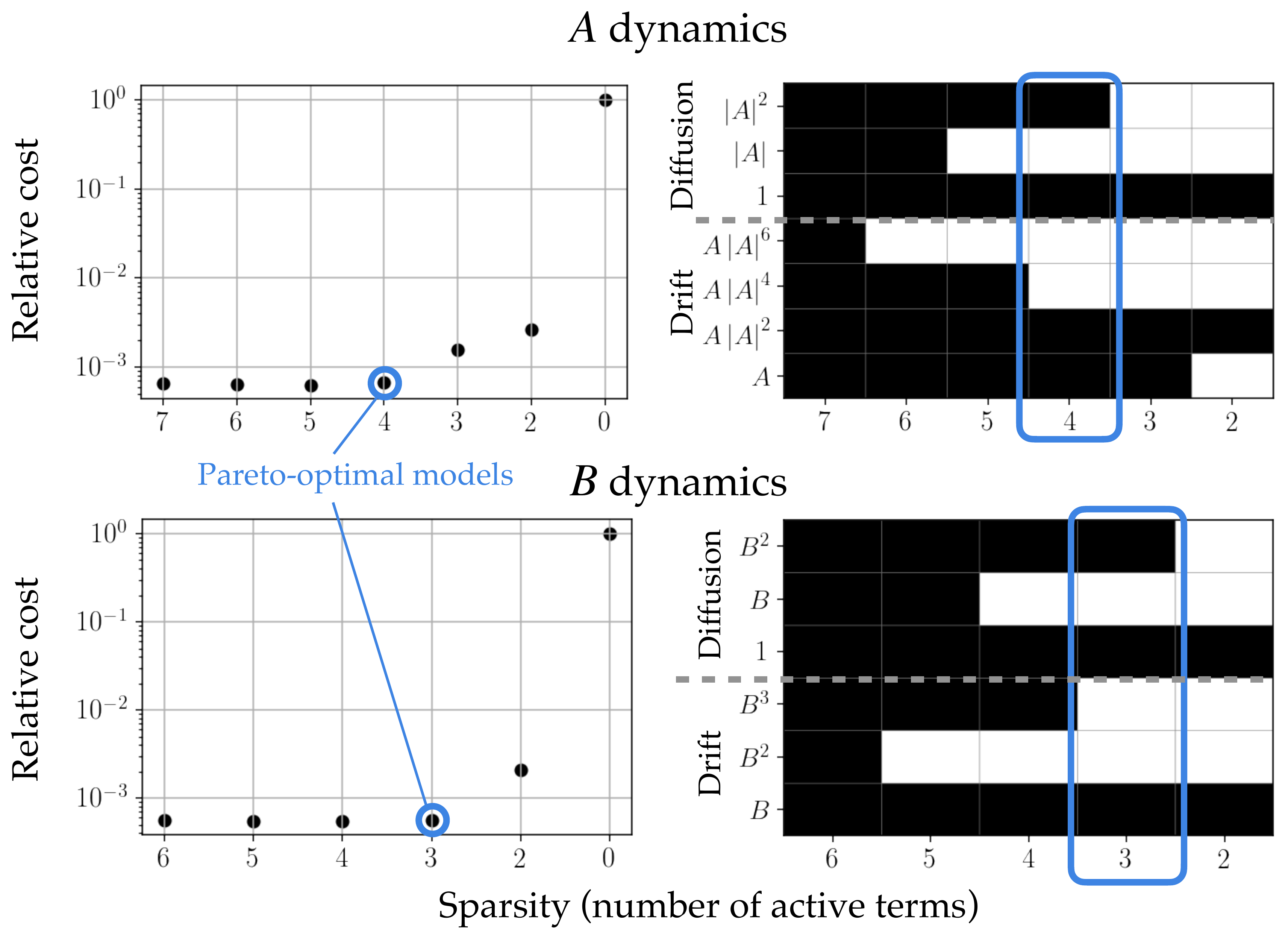}
	\end{overpic}
	\caption{Model selection with Langevin regression. The drift and diffusion functions are sparse linear combinations from a library of candidates.  We select the simplest models that are statistically consistent (small cost function).
	The vertical axes are scaled by the value of the cost function with an identically zero coefficient vector, indicated by the model with zero active terms.}
	\label{fig:ssr}
\end{figure}

The results of the stepwise sparse regression are shown in Figure~\ref{fig:ssr}.
For both the $A$ and $B$ dynamics there is a clear Pareto-optimal model.
That is, we can clearly select a model with minimal complexity in the sense that neglecting any more terms would cause the cost function to significantly increase.
The full identified model is
\begin{subequations}
\label{eq:full-dynamics}
\begin{align}
    \dv{A}{t} &= \lambda A - \mu A|A|^2 + (\sigma_A + \gamma_A|A|^2)w_1(t)\\
    \dv{B}{t} &= -\alpha B + (\sigma_B + \gamma_B B^2) w_2(t).
\end{align}
\end{subequations}

The drift function for the order parameter $A$ identified by SSR has the form of the Langevin-Stuart-Landau equation~\eqref{eq:A-dynamics}, while the drift of the axisymmetric deformation $B$ are linear relaxation dynamics.
Physically, since $B$ is defined as the difference between the instantaneous axisymmetric component of the field and that given by the conditional average on $|A|$, this models noisy relaxation towards the location on the slow manifold defined by the instantaneous value of $A$.
This is consistent with the Fourier-decomposed Navier-Stokes equations; the axisymmetric field does not instantaneously reach the equilibrium corresponding to the amplitude of the antisymmetric instability, although the relaxation timescale is often considered negligible in weakly nonlinear laminar flows~\citep{Noack2003jfm, Meliga2009jfm}.

The method identifies quadratic state-dependent diffusion for both $A$ and $B$.
Similar state-dependent diffusion arises in the case that slow macroscopic dynamics are averaged over fast degrees of freedom that are excited by stochastic forcing due to nonlinear coupling across scales~\citep{Majda2001book, Pradas2012}.
In particular, the diffusion functions in \eqref{eq:full-dynamics} have the same form as a Taylor expansion of the diffusion derived by~\citet{Pradas2012} for an unstable mode with quadratic coupling to stable modes driven by additive white noise.
The state-dependent diffusion also allows the Langevin model to better resolve the long tails of the probability distributions, as previously observed by~\citet{Callaham2021langevin}.

The identified coefficients are shown in Table~\ref{tab:coeffs}.
The close correspondence of $\lambda$ and $\mu$ reflects the fact that the order parameter is normalized so that $\bar{A} = 1$; the mean amplitude of the stochastic Stuart-Landau equation~\eqref{eq:A-dynamics} occurs near $\sqrt{\lambda/\mu}$.
Similarly, by symmetry we expect the diffusion to have equal real and imaginary parts, which is approximately true for these coefficients.

\begin{table}[tb]
\centering
 \begin{tabular}{||c | c | c | c | c | c | c||} 
 \hline
$\lambda $ & $\mu$  & $\sigma_A$ & $\gamma_A$ & $\alpha$ & $\sigma_B$ & $\gamma_B$ \\ [0.5ex] \hline
 $1.95$ & $-1.86$ & $0.80+0.70i$ & $0.27 +0.35i$ & $-26.68$ &  $6.78$ &  $0.18$ \\
 \hline
\end{tabular}
\caption{Identified coefficients for the Langevin models~\eqref{eq:full-dynamics}.}
\label{tab:coeffs}
\end{table}

\begin{figure}
	\centering
	\begin{overpic}[width=\linewidth]{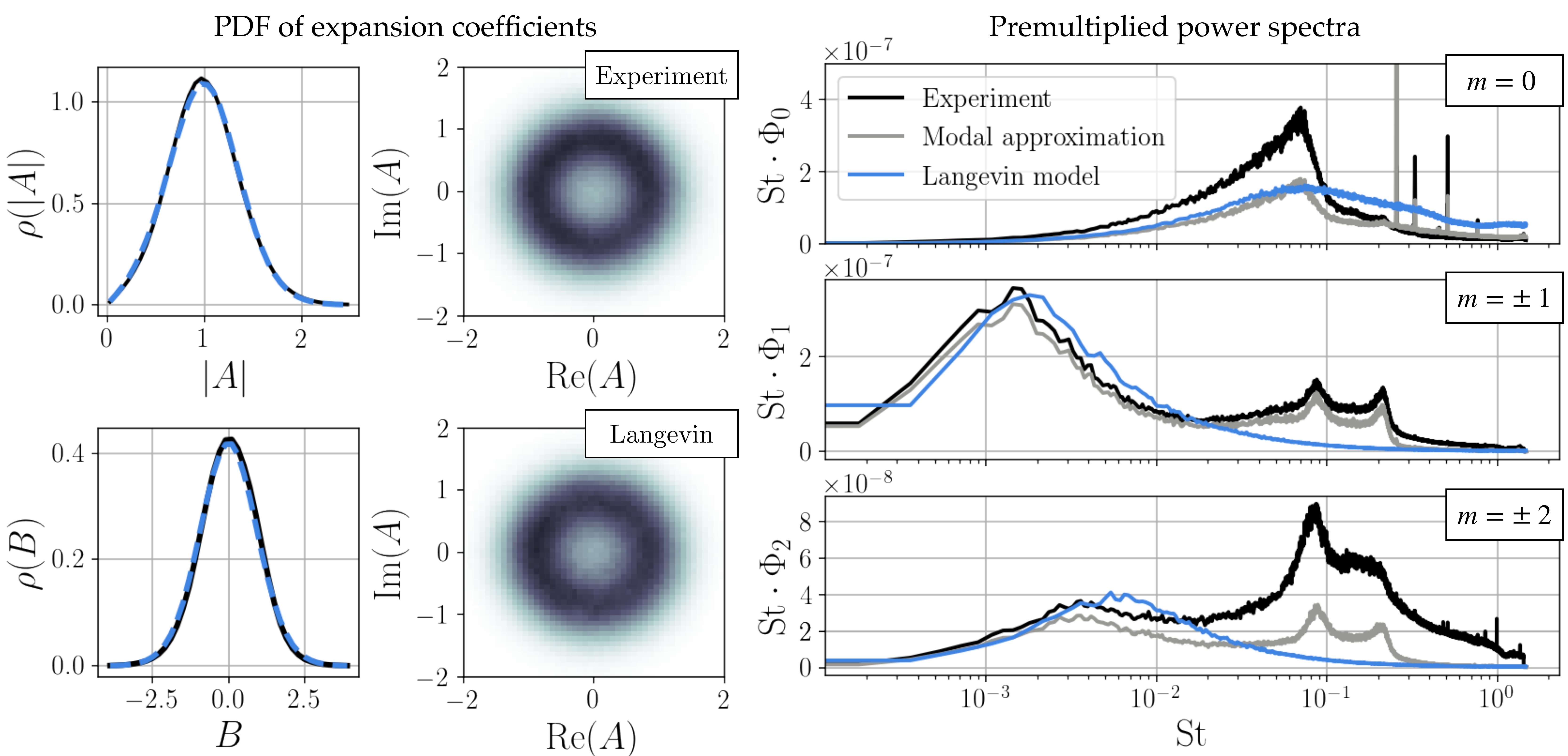}
	\end{overpic}
	\caption{Statistical evaluation of the model.
	The mean-field approximation~\eqref{eq:reconstruction} accurately captures most features of the radially averaged premultiplied power spectrum based only on the order parameters $A(t)$ and $B(t)$ (gray).
	Monte Carlo evaluation of the Langevin model (blue) shows that it reproduces the the empirical probability distributions and dominant frequency content at each azimuthal wavenumber.}
	\label{fig:results}
\end{figure}

We integrate the Langevin models with the SRIW1 stochastic Runge-Kutta method~\citep{Rossler2010sde} and construct approximate pressure fields using the parametric modal expansion~\eqref{eq:reconstruction}.
The results are compared to both the experimental data and the reconstruction based on the modal expansion and experimental values of $A$ and $B$ in Figure~\ref{fig:results} using both the empirical probability distribution and the radially integrated premultiplied power spectra.
We compare the premultiplied spectra $\St \Phi_m(\St)$, where $\Phi_m(St)$ is the estimated power spectral density of the radially integrated $m$-th Fourier component, rather than the power spectral density itself since the area under the curve of the premultiplied spectrum directly corresponds to signal energy.
This makes it better suited for comparison of the dominant energy scales of the signal.

Although the model is clearly too simplified to capture the full structure of the power spectra, it does reproduce the dominant peak for each of the leading wavenumbers and accurately approximates the probability distributions.
The compact empirical model~\eqref{eq:full-dynamics} therefore resolves the dominant physical mechanisms associated with symmetry-breaking in the turbulent wake, including linear instability, mean-field deformation, and the influence of higher harmonics.
Critically, the model reproduces the dominant frequency peak in the axisymmetric power spectrum, previously described as a ``bubble-pumping mode"~\citep{Berger1990jfs, Rigas2014jfm, Boujo2019jfs}, although it does not appear in any stability analysis.
This model suggests that this peak in the frequency spectrum may instead be viewed as the result of a finite relaxation timescale towards the axisymmetric deformation induced by the amplitude of the instability mode.
 
\section{Discussion}\label{Sec:Discussion}

We have proposed a simplified mean-field model of the symmetry-breaking behavior in a turbulent axisymmetric wake.
The empirical model comprises seven parametrically varying spatial modes and three dynamical degrees of freedom, and was constructed entirely from experimental observations.
Using a phase-aligned conditional average with respect to the order parameter, we have shown that the fixed-mode ansatz of standard modal decompositions cannot explain the mean-field deformation related to the symmetry-breaking instability of the turbulent axisymmetric wake.
This reflects the higher-order influences as well as nonlinear self-interaction of the axisymmetric part of the flow, both of which are typically negligible for weakly supercritical laminar flows.
Modeling approaches based on weakly nonlinear approximations have proven highly successful in laminar flows.
However, in this work we have shown that the extension of this perspective to turbulent flows is more subtle than adding stochastic forcing to the weakly nonlinear model.
The amplitude scaling and structure of the axisymmetric deformation is inconsistent with the polynomial scaling implied by such an approach, although the conditional averaging can capture the natural behavior of the mean flow deformation.

Furthermore, we find that the modal decomposition based on this conditional average does not explain the peak in the axisymmetric part of the power spectrum previously identified as ``bubble-pumping''.
However, since this modulation of the recirculation bubble is not related to any known linear stability mode we address this apparent discrepancy by introducing a finite relaxation time of the axisymmetric part of the field to that predicted by the instantaneous order parameter.
The spatial field associated with these dynamics is the gradient of the manifold defined by the conditional average.
These relaxation dynamics are consistent with the underlying Navier-Stokes equations, but this timescale is often considered negligible for weakly nonlinear laminar flows.

We evaluate the proposed three-dimensional, seven-mode stochastic mean-field model by fitting it to the experimental data using the Langevin regression nonlinear model identification framework.
This method recovers the leading-order deterministic terms of a qualitative analogy to Landau's theory of symmetry-breaking phase transitions, as well as a quadratic nonlinear state-dependent noise model.
This form of diffusion is consistent with analysis of fast-slow systems with quadratic nonlinearities where only the fast scales are stochastically forced~\citep{Majda2001cpam, Pradas2012}.
Monte Carlo evaluation of the model matches the stationary probability distributions of the experimental data and resolves the dominant peak in the power spectrum at the leading azimuthal wavenumbers.

Beyond the context of the axisymmetric wake, these results  support the parameterization of turbulent fluctuations as stochastic forcing of the quasi-deterministic coherent structures evolving near a slow manifold, at least as an approximation for empirical models.
We emphasize that this description relies on a strict separation of scales, which is known to be absent in turbulent flows.
Even with this caveat, this simplification is appealing enough for engineering applications such as closed-loop flow control that it may be useful even if it only holds in an approximate sense.
For example, a feedback controller based on a nonlinear Langevin model has been shown to produce power-efficient drag reduction on an Ahmed body with a similar symmetry-breaking instability~\citep{Brackston2016jfm}. 
By approximating turbulent fluctuations as process noise, Langevin models are a natural starting point for controller design.
Moreover, least-order models such as the one proposed in this work may be compact enough to be used in real-time nonlinear stochastic optimal control schemes.

More broadly, we hope that this work illustrates a principled approach to constructing stochastic reduced-order models from limited experimental observations of a turbulent flow.
Although much progress has recently been made in developing stochastic models of turbulent flows using the linearized governing equations~\citep{Zare2017jfm, Towne2021aiaa} and with empirical models of experimental data~\citep{Rigas2015jfm, Sieber2020, Herrmann2020}, there are many open questions about how the heuristics of low-dimensional models of weakly nonlinear flows will translate to fully-developed turbulence.
In this work we have chiefly focused on the mean-field deformation associated with the symmetry-breaking bifurcation, but recent studies have highlighted the role of higher-order triadic nonlinear interactions in capturing the dynamics of both natural~\citep{Schmidt2020bispectral} and actuated~\citep{Herrmann2020} turbulent wakes.
Capturing the interactions between instability modes (for instance, symmetry-breaking and vortex-shedding) may also prove critical in developing models suitable for active flow control.
Spectral~\citep{Schmidt2018jfm, Towne2018jfm} and bispectral~\citep{Schmidt2020bispectral} decompositions are particularly promising avenues for identifying space-time coherent structures from observations of turbulence.

As fully empirical data-driven methods continue to grow in popularity and utility, ensuring consistency with the underlying physics remains a central challenge.
Although the model herein relies on an unphysical approximation of near-equilibrium thermodynamics, it nevertheless captures many of the essential statistical features of the data and leads to new hypotheses about the behavior of the axisymmetric wake in particular and globally unstable turbulent flows in general.

\section*{Acknowledgements} SLB acknowledges funding support from the Army Research Office (ARO W911NF-19-1-0045; program manager Dr. Matthew Munson).
JLC acknowledges funding support from the Department of Defense (DoD) through the National Defense Science \& Engineering Graduate (NDSEG) Fellowship Program. GR acknowledges funding support from the Engineering and Physical Sciences Research
Council (grant number EP/I005684) for the experimental data.

\clearpage
\setlength{\bibsep}{2.4pt plus 1ex}
\begin{spacing}{.01}
	\small
	\bibliographystyle{jfm}
	\bibliography{refs}
\end{spacing}

\end{document}